\documentclass[utf8
               ]{jetp} 
\usepackage{epstopdf}
\usepackage{subfig}

\begin{document}
\English

\title{Kohn anomalies in momentum dependence of magnetic susceptibility of some three-dimensional systems}




\email{katanin@mail.ru} 



\setaffiliation1{Ural Federal University, Mira str. 19, 620002, Ekaterinburg, Russia}
\setaffiliation2{M. N. Mikheev Institute of Metal Physics UrB RAS, Kovalevskaya str., 18, 620990, Ekaterinburg, Russia} 
\setauthor{A.~A.}{Stepanenko}{{1}}
\setauthor{D.~O.}{Volkova}{{1}}
\setauthor{P.~A.}{Igoshev}{{2}{1}}  
\setauthor{A.~A.}{Katanin}{{2}{1}} 



\abstract{
We study a question of presence of Kohn points, yielding at low temperatures non-analytic momentum dependence of magnetic susceptibility near its maximum, in electronic spectum of some three-dimensional systems. In particular, we consider one-band model on face centered cubic lattice with hopping between nearest and next-nearest neighbors, which models some aspects of the dispersion of ZrZn$_2$, and the two-band model on body centered cubic lattice, modeling the dispersion of chromium. For the former model it is shown that Kohn points yielding maxima of susceptibility exist in a certain (sufficiently wide) region of electronic concentrations; the dependence of the wave vectors, corresponding to the maxima, on the chemical potential is investigated. For the two-band model we show existence of the lines of Kohn points, yielding maximum of the susceptibility, which position agrees with the results of band structure calculations and experimental data on the wave vector of antiferromagnetism of chromium.}

\maketitle


\section{Introduction}

Pecularities of an electronic spectrum have strong impact on physical properties of strongly-correlated systems. In particular, peaks of the density of states (occuring, e.g., due to van Hove singularities) lead to ferromagnetism,  while nesting of the Fermi surface results in a tendency to spin density wave formation. Although an ``ideal'' nesting is rather rare (especially in real magnetic substances), spin density waves can be caused by ``local'' nesting, which is present near certain points of Fermi surface, connected by a spin density wave vector $ {\bf Q} $ and having opposite Fermi velocities.   

Study of the effect of this ``local'' nesting originates from W. Kohn's pioner paper \cite {Kohn}, which has shown that the susceptibility of systems with spherical Fermi surface of radius $k_F $ depends non-analytically on momentum close to wave vectors of the length $Q=2k_F$, connecting points on the opposite sides of the Fermi surface, which have also opposite velocities (so-called Kohn points). Generally this non-analytical peculiarity is rather weak, leading, however, to a number of important phenomena, such as anomalies of phonon frequencies, Kohn-Luttinger mechanism of superconductivity  \cite {KohnLuttinger}, etc. 

The strongest effect of Kohn anomalies of susceptibility on magnetic properties can be expected in the case when the wave vector $ \mathrm {\mathbf {Q}} $, connecting Kohn points, corresponds to a global (non-analytical) maximum of the susceptibility, such that $ \mathrm {\mathbf {Q}} $ coincides with a wave vector of spin- or charge density wave.  While the property of a maximum of susceptibility being global depends on the whole electronic structure, and can be established, as a rule, only numerically or experimentally, the presence of a local maximum of susceptibility follows from local geometry of Fermi surface near Kohn points. In particular, L. M. Roth with co-authors \cite {Roth}, and also T. M. Rice \cite {Rice} have established that in one-band models the susceptibility has a local maximum at the wave vector $ \mathrm {\mathbf {Q}} $ if the Fermi surface near each of the Kohn points, separated by the vector $ {\bf Q} $, has opposite curvature in two perpendicular directions, or curvature in one of the directions  is equal to zero. 
The latter condition, in particular, is fulfilled for two-dimensional Fermi surfaces; the case of circular Fermi surface was explicitly studied by Stern \cite {Stern}, later T. Holder and W. Metzner \cite {Holder2012, Metzner} have considered the effect of Kohn points for arbitrary two-dimensional 
Fermi surfaces, in particular they have investigated peculiarities of spin and charge susceptibilities, as well as self-energy of electrons.

Although at finite temperatures $T$ non-analytical behaviour of susceptibility, caused by Kohn points, takes place only at momenta $ | {\bf q-Q} | \gtrsim T/v_F $ ($v_F $ is the Fermi velocity at Kohn points), it becomes important in the vicinity of quantum phase transitions at $T\rightarrow 0$. Recently, T. Sch\"afer, et al. \cite {Katanin} investigated effect of Kohn points on quantum critical behaviour and have shown that corresponding critical exponents can strongly deviate from predictions of Hertz-Moriya-Millis  theory because of the presence of Kohn anomalies. In that study, however, only simple cubic lattice with nearest neighbour hopping, for which there are lines of Kohn points, has been investigated.

Depending on momentum dependence of susceptibility, two different cases, in which one can expect an essential effect of Kohn points near quantum phase transitions, can be distinguished. In the first case  the vector of a spin density wave is small, and the competition of ferromagnetism and spin density  wave takes place. This situation is possibly realised near quantum phase transition in ZrZn$ _ 2$ \cite {ZrZn2_QPT1,ZrZn2_QPT2,ZrZn2_QPT3,ZrZn2_QPT4}, where sharp change of a critical exponent of electrical resistivity under pressure is observed \cite {ZrZn2_rho}. 
In the second case the wave vector of spin density is not small (as it happens, e.g., in chromium \cite {Cr}), and gradual suppression of Neel temperature of an incommensurate order by external factors, e.g. pressure \cite {Cr_QPT} or doping \cite {Cr_dop1, Cr_dop2, Cr_dop3} takes place. 

The aim of the present paper is to consider electronic systems on various three-dimensional lattices, which lead to Kohn anomalies, and model the two above described types of momentum dependence of susceptibility. In particular, consideration of face-centered cubic (fcc) lattice with hopping between nearest- and next-nearest neighbours (Section 2 of the paper) allows to describe the situation with a small wave vector (caused by contribution of Kohn points), as well as its gradual evolution to a spin density wave vector, essentially different from zero. Although evolution of a wave vector of an incommensurate order with changing concentration of electrons on fcc lattice was investigated earlier \cite {Igoshev, Igoshev1}, the relation of the obtained wave vectors to the geometry of a Fermi surface, and also the effect of Kohn points on magnetic susceptibility was not considered. To model the Fermi-surfaces of chromium,
the two-band dispersion on body-centered cubic (bcc) lattice is also considered (see Section 3 of the paper). For each of the these cases Kohn points are determined and their contribution to a static magnetic susceptibility is calculated.

\section{The one-band model}

\subsection{Formulation of the model, corresponding Fermi surfaces and susceptibility}

We consider the Hamiltonian of the Hubbard model
\begin{equation}
\label{eq1}\hat H=-\sum_{i,j,\sigma}t_{i,j}\hat c_{i\sigma}^\dag\hat c_{j\sigma}+U\sum_{i}\hat n_{i\uparrow}\hat n_{i\downarrow},
\end{equation}
where $\hat c_{i\sigma}^\dag$($\hat c_{i\sigma}$) are the Fermi-operators of creation (annihilation) of an electron at site $i$ with spin projection $\sigma=\uparrow,\downarrow$, $\hat n_{i\sigma}=\hat c_{i\sigma}^\dag\hat c_{i\sigma}$ is the operator of number of electrons at site $i$, $t_{i,j}$ is the hopping parameter, $U$ characterizes the on-site repulsion of electrons and determines the strength of electron correlations.
To calculate magnetic susceptibility, we use generalized random phase approximation (RPA), see e.g., Ref. \cite{Tremblay},
\begin{equation}\label{eq2}
\chi_{\mathrm{\mathbf{q}}}=\frac{\chi_{\mathrm{\mathbf{q}}}^{0}}{1-{U_{\rm eff}}\chi_{\mathrm{\mathbf{q}}}^{0}},
\end{equation}
where
\begin{equation}\label{chi0}
\chi_{\mathrm{\mathbf{q}}}^{0}=-\sum_{\mathrm{\mathbf{k}}}\frac {f_{\mathrm{\mathbf{k}}}-f_{\mathrm{\mathbf{k+q}}}}{E_{\mathrm{\mathbf{k}}}-E_{\mathrm{\mathbf{k+q}}}}
\end{equation}
is the susceptibility of the system of non-interacting electrons,  $E_{\mathrm{\mathbf{k}}}=\sum_\delta t_{i,i+\delta} \exp(i \mathrm{\mathbf{k}} \mathrm{\mathbf{R}}_\delta)$ is the dispersion, $f_{\mathrm{\mathbf{k}}}\equiv f(E_{\mathrm{\mathbf{k}}})$ is the Fermi function,  $\bf q$ is the wave vector,  $\bf k$ is the electronic quasimomentum, $U_{\rm eff}$ is the value of an effective inter-electron interaction in the particle-hole spin channel. This approximation (with some parameter $U_{\rm eff}<U$) is qualitatively applicable for the considered three-dimensional systems due to sufficiently weak momentum dependence of the self-energy and vertices of electron-electron interaction, irreducible in the particle-hole channel \cite{Tremblay}.  
The value
$U_{\rm eff}$, which can be determined, e.g., from the sum rules, 
does not play an important role for further consideration; it is assumed only, that it is sufficient to form (weakly) magnetically ordered state, i.e. $U_{\rm eff} \chi^0_{\bf Q}\approx 1$. The applicability of qualitative results of RPA for studying Kohn anomalies in strongly correlated three-dimensional systems was shown, e.g., in Ref. \cite{Katanin}.

Below we consider fcc lattice with hoppings between nearest neighbors $t$ and next nearest neighbors $-t'$, corresponding to the dispersion
\begin{eqnarray}
E_\mathrm{\mathbf{k}}&=&-4t\left(\cos{\frac {k_x}{2}}\cos{\frac {k_y}{2}}+\cos{\frac {k_x}{2}}\cos{\frac {k_z}{2}}+\cos{\frac {k_y}{2}}\cos{\frac {k_z}{2}}\right)+2t'\left(\cos{k_x}+\cos{k_y}+\cos{k_z}\right)-\mu,
\end{eqnarray}
where $\mu$ is the chemical potential, the lattice constant is set equal to unity.
\begin{figure}
\centering
\subfloat[]{
\label{fig3a}
\includegraphics[width=0.3\textwidth]{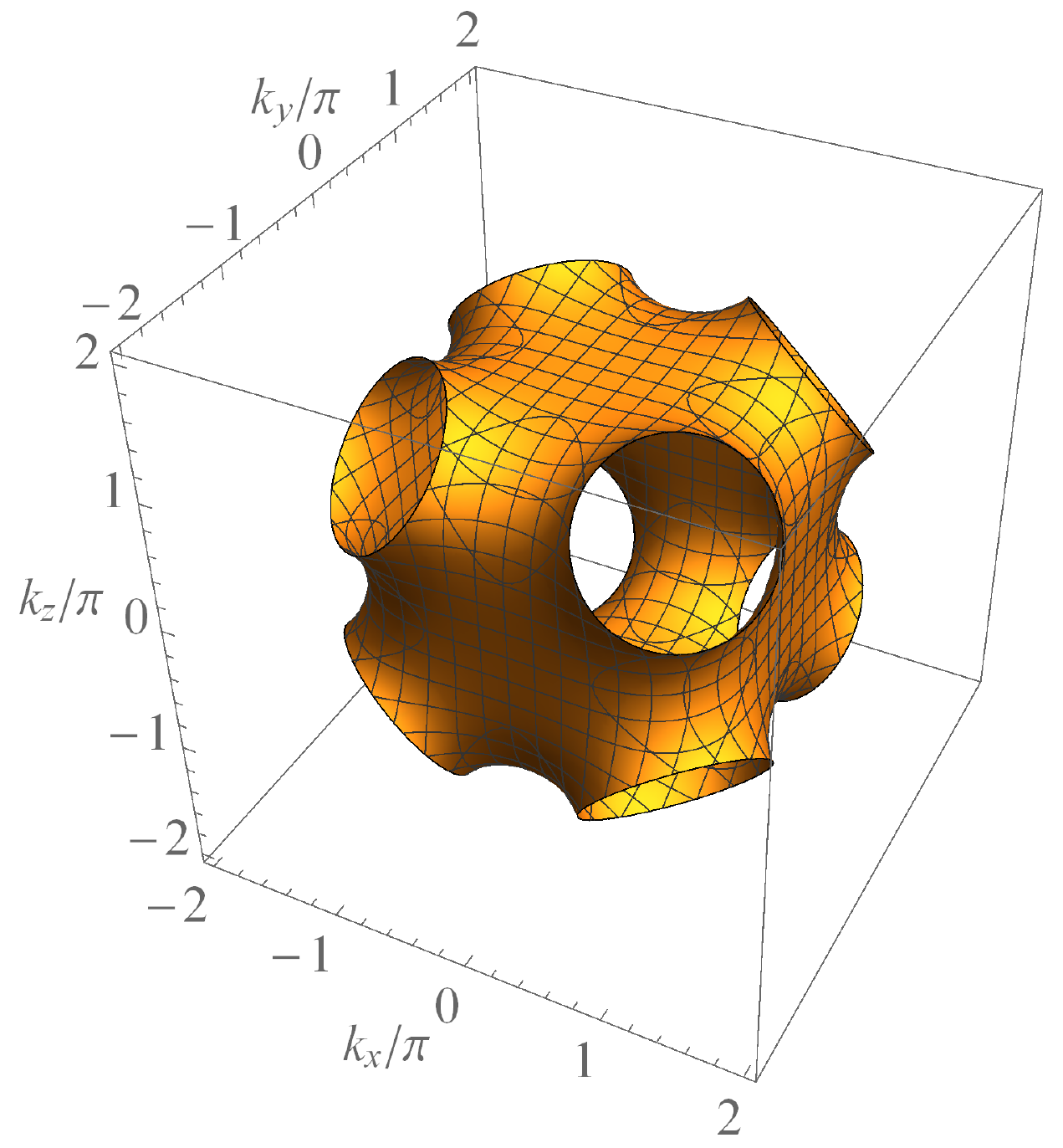}}
\subfloat[]{
\label{fig3b}
\includegraphics[width=0.3\textwidth]{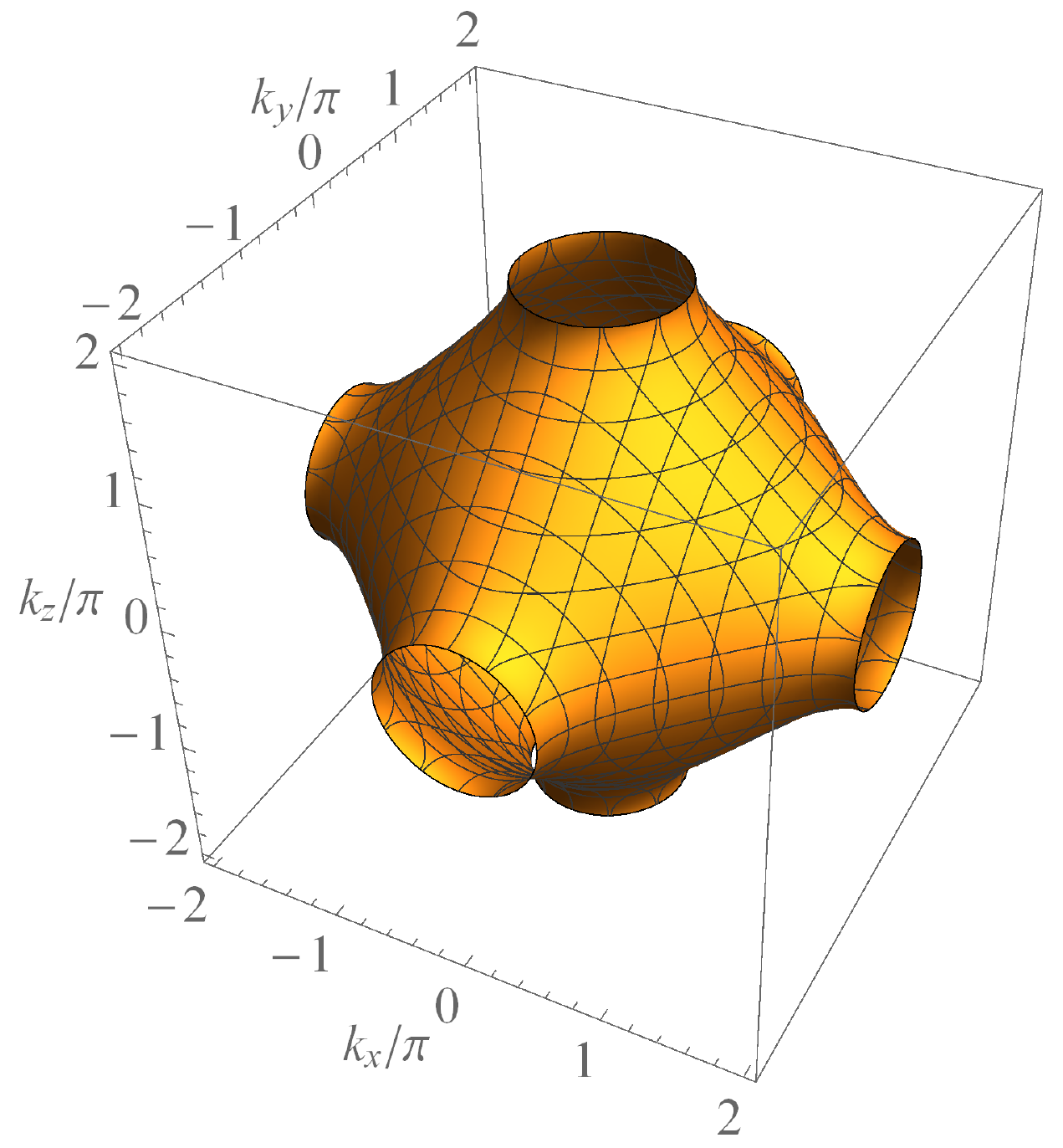}}
\subfloat[]{
\label{fig3c}
\includegraphics[width=0.3\textwidth]{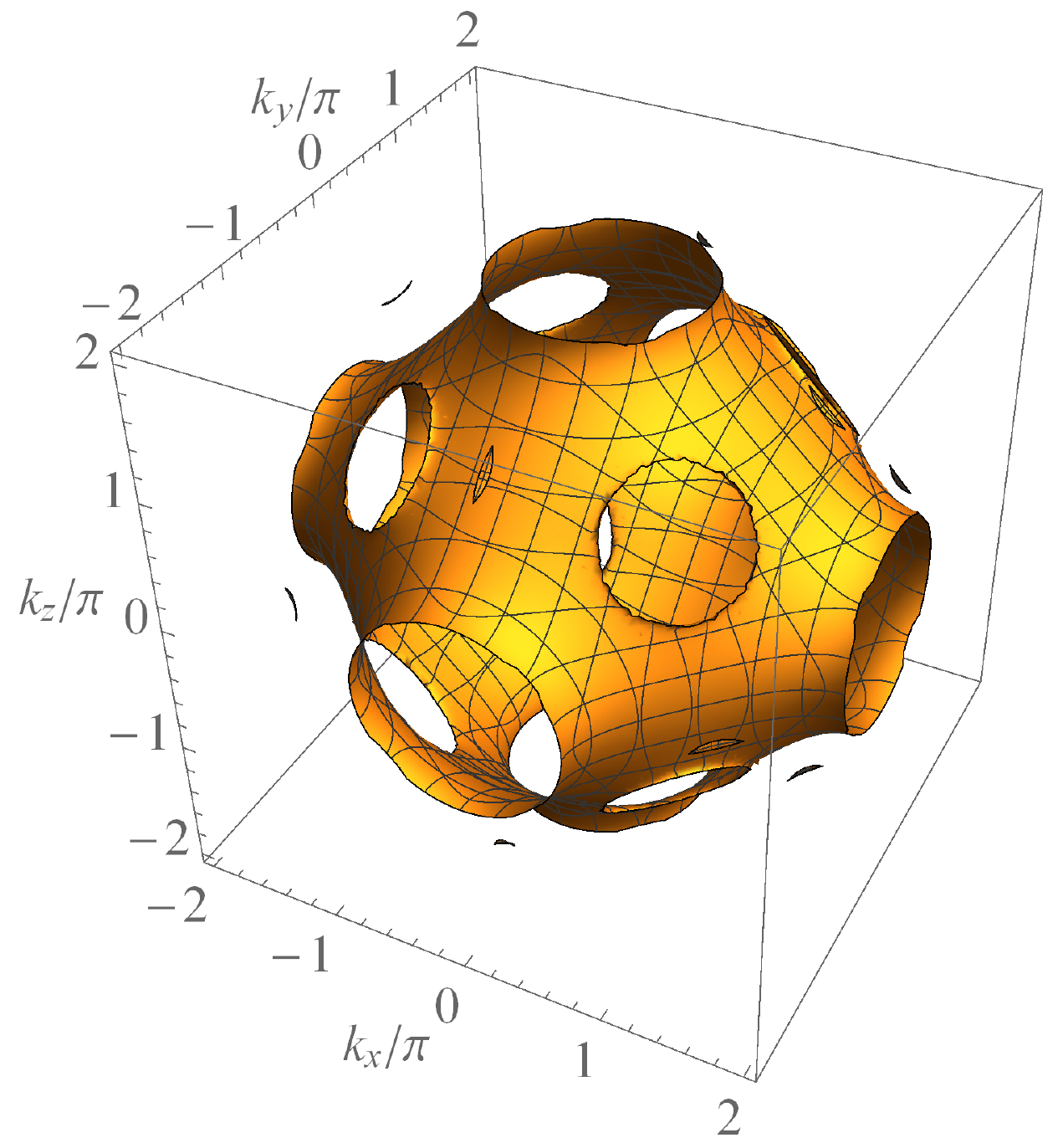}}
\caption{\label{figFS}The Fermi surfaces of electrons on fcc lattice with hopping between nearest and the next nearest neighbors (a) $t'=0.3t$, $\mu=1.0t$ (b) $t'=-0.45t$, $\mu=2.4t$ (c) $t'=-0.45t$, $\mu=2.8t$}
\end{figure} 
Depending on the values of $t'/t$ and $\mu/t$, the Fermi surface can be either connected (for $\mu$ not too close to the upper edge of the band), or consist of several disconnected parts. We consider the former case, as the most physically interesting, examples of the Fermi surfaces for some values $t'/t$, $\mu/t$ are shown in Fig.~\ref{figFS}. 

\begin{figure}[h]
\centering
\subfloat[]{
\includegraphics[width=0.3\textwidth]{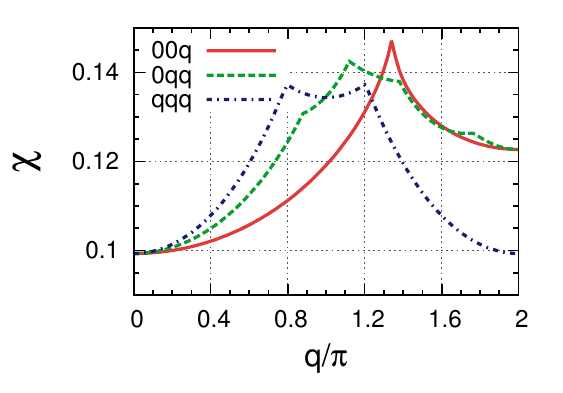}}
 \subfloat[]{
\includegraphics[width=0.3\textwidth]{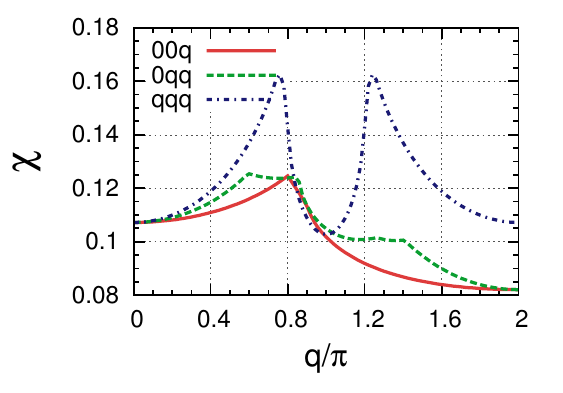}}
\subfloat[]{
\includegraphics[width=0.3\textwidth]{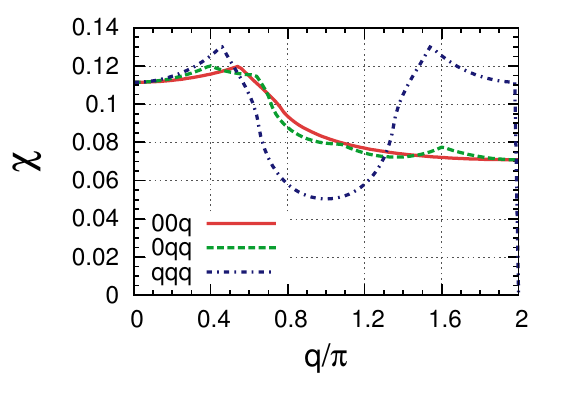}}
  \caption{The momentum dependences of magnetic susceptibility $\chi_{\bf q}^0$ of non-interacting electrons 
 at $T=0$ on fcc lattice ($t'=0.3t$) along symmetric directions: dot-dashed lines correspond to the direction $(q,q,q)$; dotted lines -- to the direction $(0,q,q)$; solid lines  -- to the direction $(0,0,q)$, at different values of the chemical potential: (a) $\mu=0.8t$;
(b) $\mu=3.3t$;
(c) $\mu=4t$
}\label{fig5}

\end{figure}

\begin{figure}[h]
\centering
\subfloat[]{
\includegraphics[width=0.3\textwidth]{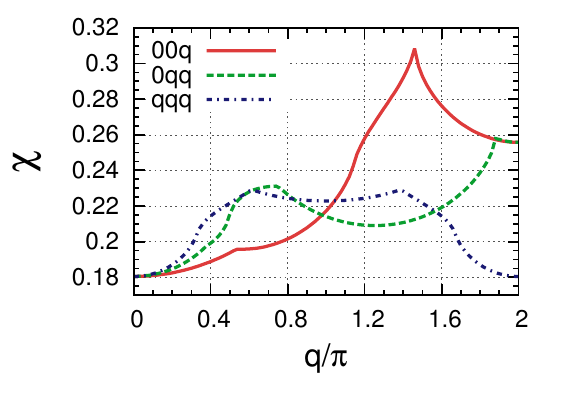}}
\subfloat[]{
\includegraphics[width=0.3\textwidth]{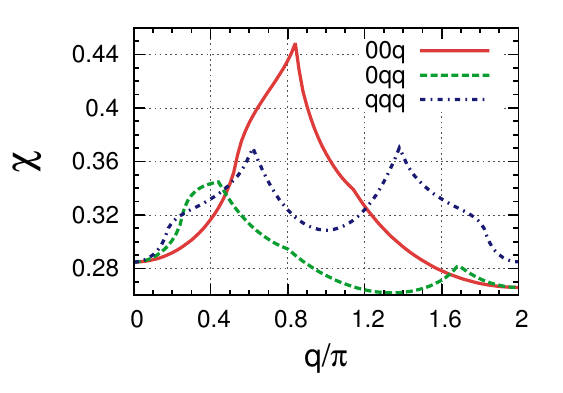}}
\subfloat[]{
\includegraphics[width=0.3\textwidth]{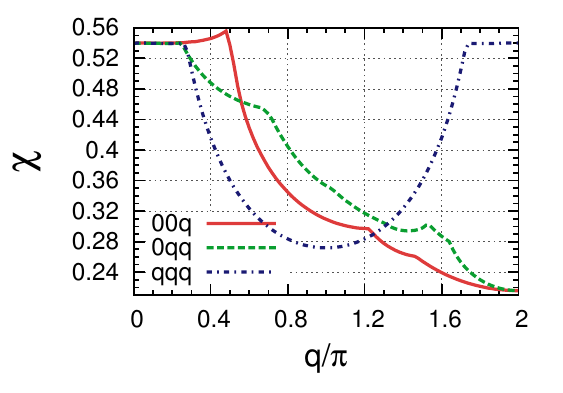}}\\
\caption{The same as in Fig. \ref{fig5} at $t'=-0.45$ and chemical potentials (a) $\mu=1.6t$; (b) $\mu=2.4t$; (c) $\mu=2.8t$. }
\label{fig6}

\end{figure}

The momentum dependences of bare susceptibilities $\chi_{\bf q}^0$, determined numerically from the equation (\ref{chi0}) for some values of the parameters $t'/t$ and $\mu$ at the temperature $T=0$ are shown in Figs. \ref{fig5},\ref{fig6}. It can be seen that these susceptibilities have nonanalytic dependences near maxima and inflection points, which connection with the Kohn points is discussed in the following subsections.

\subsection{The contribution of Kohn points to the non-uniform susceptibility}

\label{Sec:Kohn_points}

In general case Kohn points ${\mathbf K}$ and $\mathbf{K+Q}$, leading to a nonanalytic momentum dependence of the susceptibility correspond to the points of the Fermi surface ($E_{\bf K}$=$E_{\bf K+Q}=0$), having opposite Fermi velocities.
As L. M. Roth with coauthors \cite{Roth} and T. M. Rice \cite{Rice}  have shown, under certain conditions on the curvature of the Fermi surface, the magnetic susceptibility can have a nonanalytic maximum on the wave vector, connecting two Kohn points. In order to formulate these conditions, we consider the expansion of the dispersion near the corresponding points by representing $\mathrm{\mathbf{k}}=\mathrm{\mathbf{K+k}_1}$. Introducing a local coordinate system in momentum space $\mathrm{\mathbf{k}_1}=(k'_x,k'_y,k'_z)$, which axis $k'_z$ is aligned along the Fermi velocity
$\mathrm{\mathbf{v}_{\bf K}}=(\nabla E_{\mathrm{\mathbf{k}}})_{\mathrm{\mathbf{k}}=\mathrm{\mathbf{K}}}$, and the axes $k'_{x,y}$ are rotated around the new axis $k'_z$ such that
${\partial^2E_{\mathrm{\mathbf{k}}}}/({\partial k'_x\partial k'_y})=0$, we can express the dispersion in the form
\cite{Roth,Rice}
\begin{equation}
\begin{gathered}
E_{\mathrm{\mathbf{K+k}}_1}\simeq vk'_z+\frac {(k'_x)^2}{2m_{x}}+\frac {(k'_y)^2}{2m_{y}},\\
E_{\mathrm{\mathbf{K+Q+k}}_1}\simeq-vk'_z+\frac {(k'_x)^2}{2m_{x}}+\frac {(k'_y)^2}{2m_{y}},
\end{gathered}
\label{Eexp}
\end{equation}
where $v=|{\bf v}_{\bf K}|$ and we assume that the Kohn points are connected by certain symmetry operations of the crystal lattice, so that the masses $m_{x,y}$ are the same in the first and second lines of Eq. (\ref{Eexp}). According to the results of Refs. \cite{Roth,Rice}, the magnetic susceptibility will have a (local) maximum at the wave vector $\mathrm{\mathbf{Q}}$, if one of the following conditions is fulfilled: $m^{-1}_x m^{-1}_y<0$ (so-called hyperbolic Kohn points) or
$m_x^{-1} m_y^{-1}=0$ (cylindrical Kohn points). Physically, the most interesting case corresponds to Kohn points, spaced by a wave vector along one of the symmetric directions (since only these directions can provide a global maximum of susceptibility).

In the case of hyperbolic Kohn points, the momentum dependence of magnetic susceptibility at $T=0$ near the maximum has the form \cite{Rice}
\begin{equation}
\chi_{\mathrm{\mathbf{q+Q}}}^{0}=\chi_{\mathrm{\mathbf{Q}}}^{0}-\frac{\sqrt{|m_x m_y|}}{16\pi}|q_z|, 
\end{equation}
where $q_z$ is the projection of the vector $\bf q$ on the axis $k'_z$.

Let us consider, for example, the expansion of the dispersion on fcc lattice near Kohn points, which are located symmetrically with respect to the point $X=(0,0,2\pi)$ and characterized by different directions of the spin-density wave vector $\mathbf Q$.
In this case the Fermi surface can either be ``closed'' around the point $\Gamma=0$, see Fig. \ref{figFS}a 
(the point $X$ is located outside the Fermi surface and $E_X>0$),
 or have a ``window'' near the point $X$ (see Fig. \ref{figFS}b, c, $E_X<0$).
In both cases we assume that the points $\{\mathbf{K},\mathbf{K+Q}\}=X\pm{\mathbf Q}/2$ belong to the Fermi surface.
For the sake of simplicity we will assume that $|t'|<t/2$.
We consider below different symmetric vectors $\bf Q$ that connect the Kohn points near the point $X$.

\begin{figure}
\centering
\subfloat[]{
\includegraphics[width=0.3\textwidth]{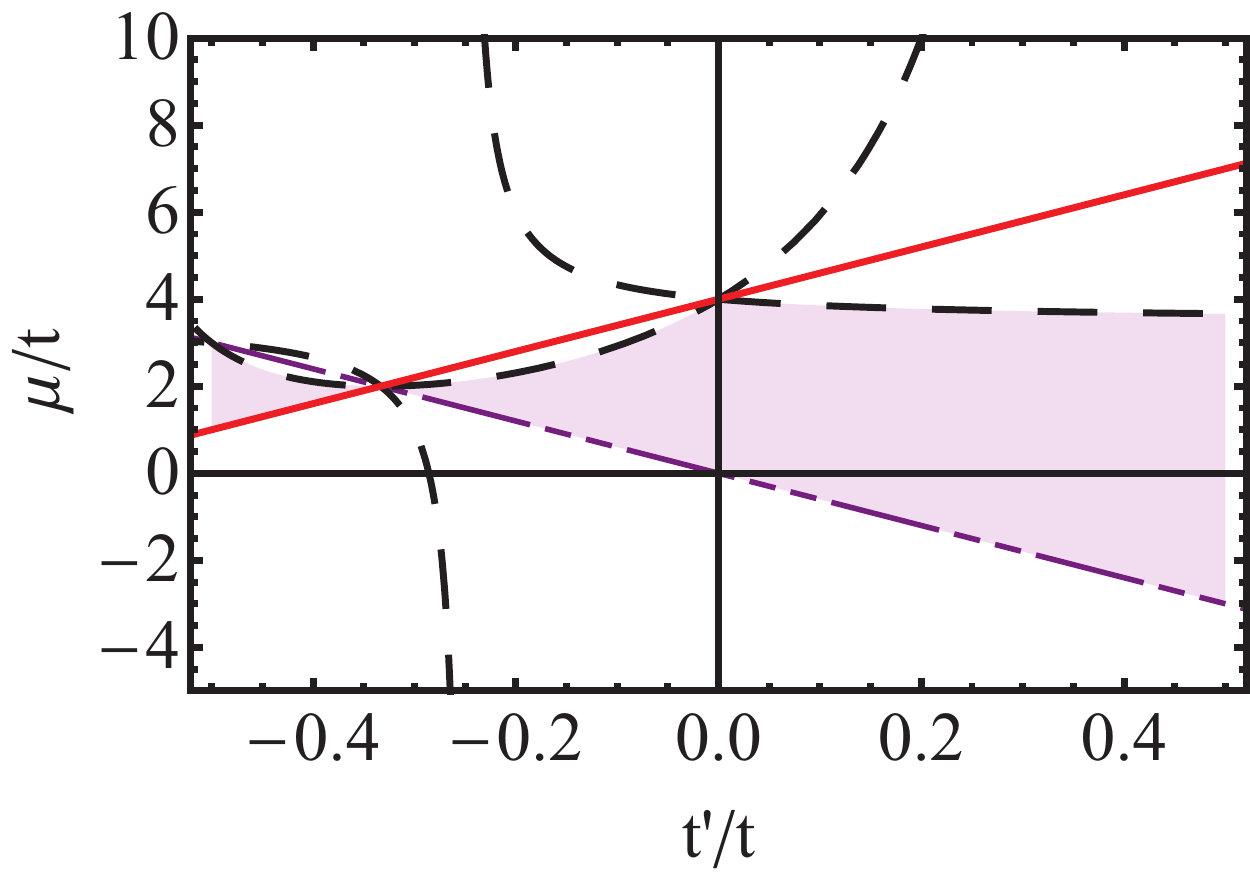}}
\subfloat[]{
\includegraphics[width=0.3\textwidth]{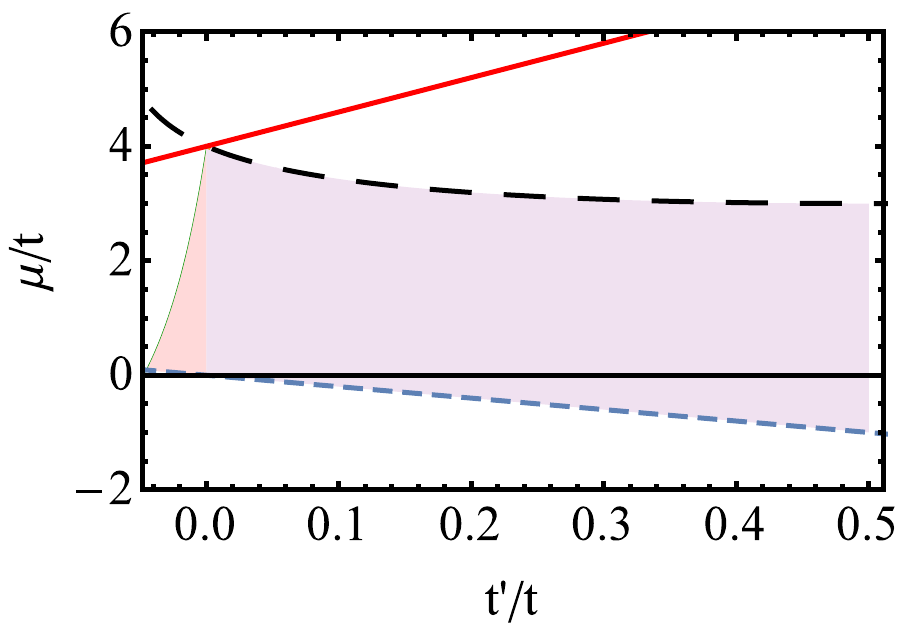}}\\
\subfloat[]{
\includegraphics[width=0.3\textwidth]{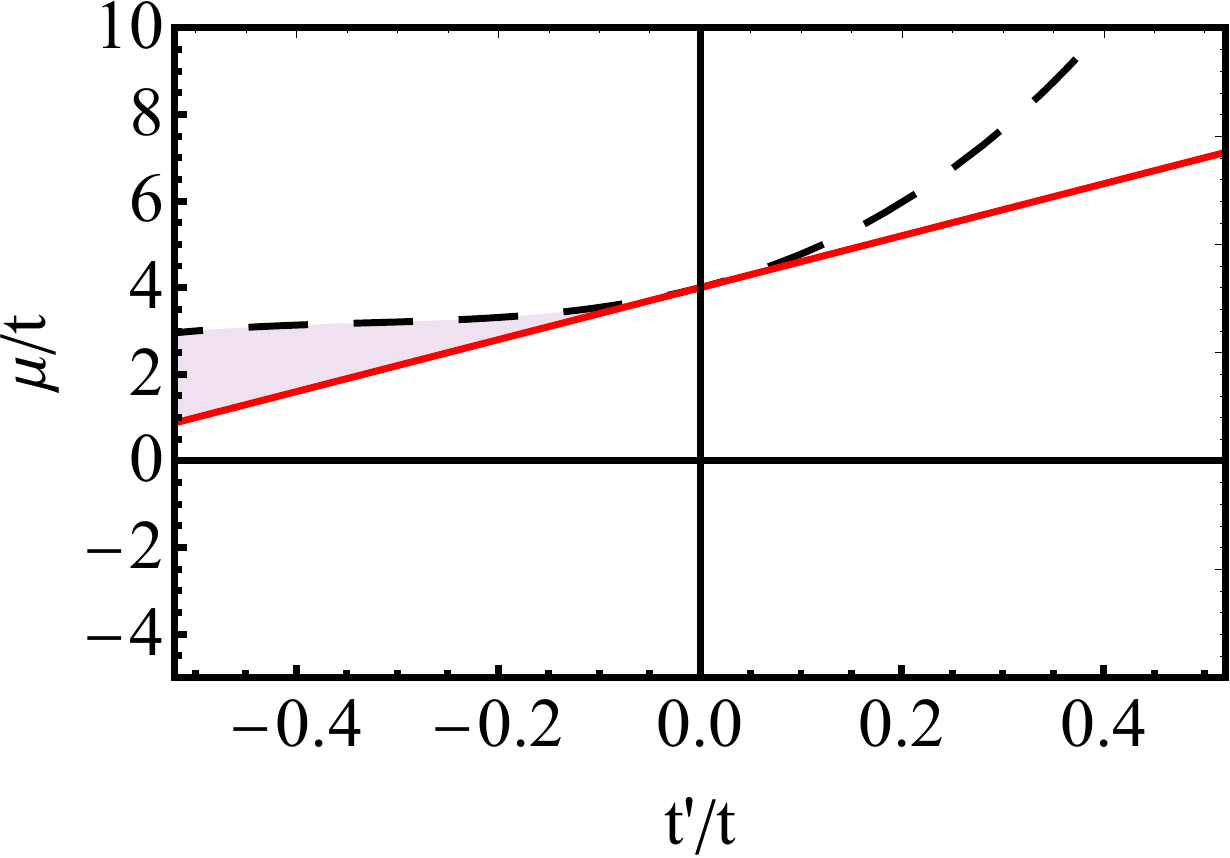}}
\subfloat[]{
\includegraphics[width=0.3\textwidth]{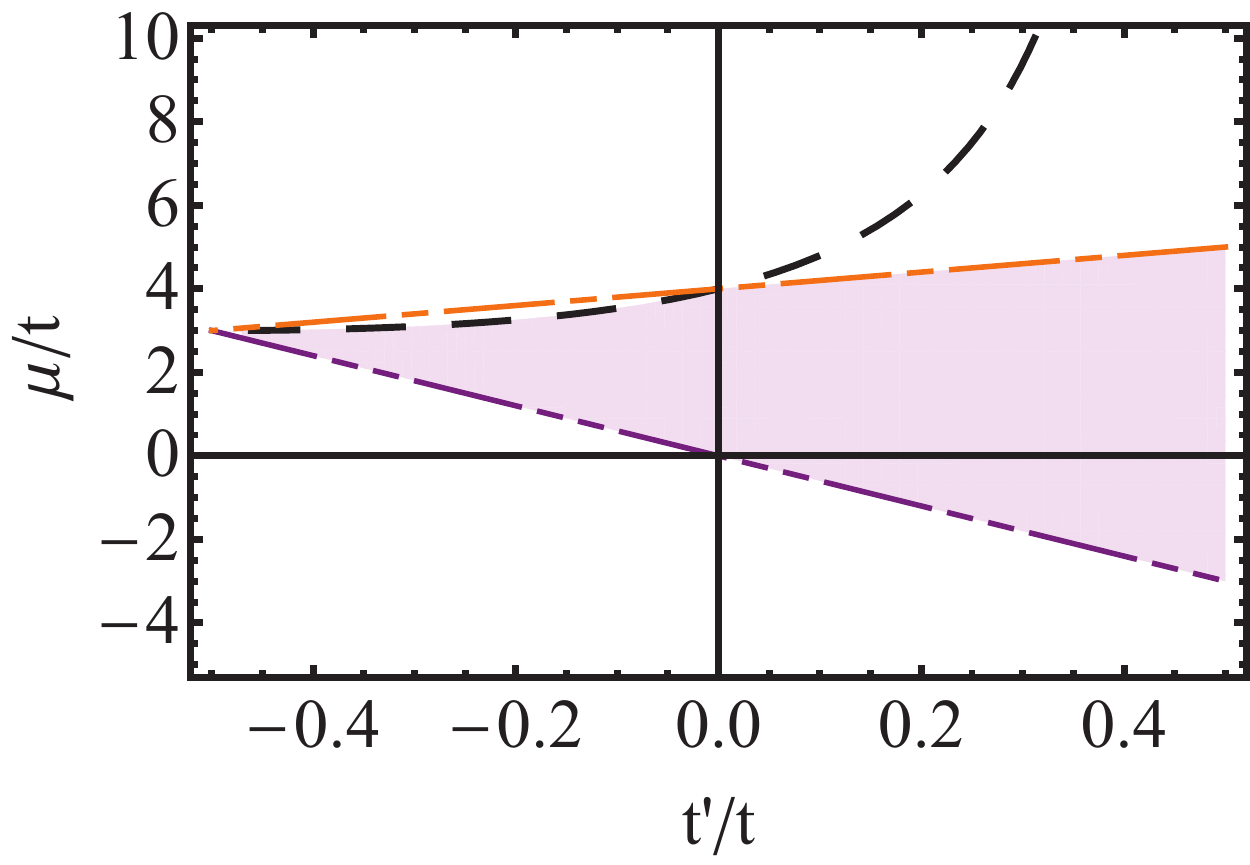}}
\caption{
The conditions for the presence of Kohn anomalies in the momentum dependence of the susceptibility for different $t'$ and $\mu$ (shaded regions) for the symmetric directions $(Q_{1X},Q_{1X},Q_{1X})$ (a), $(0,Q_{2X},Q_{2X})$ (b), $(Q_{4X},0,0)$ (c), 
connecting points of the Fermi surface in a neighborhood of the point $X$ and the direction $(Q_{4L},-Q_{4L},0)$ (d) in a neighborhood of the point $L$.
Solid lines correspond to the intersection of the point $X$ by the Fermi surface ($E_X=0$), dot-dashed lines correspond to the intersection of the point $W$ by the Fermi surface ($E_W=0$), the dotted line corresponds to the intersection of the point $L$ by the Fermi surface ($E_L=0$),  dashed lines are determined from the conditions for the equality of one of the inverse masses to zero $m^{-1}_{x,y}=0$, the solid thin line in Fig. b is determined from the condition of superiority of one mass of another 10 times, short-dashed line is the boundary of the region, where Kohn points connected by the wave vector  $(0,Q_{2X},Q_{2X})$ exist. }\label{fig_reg}
\end{figure}

1) For the direction ${\mathbf Q}=(Q_{1X},Q_{1X},Q_{1X})$, the condition $E_{\bf K}=E_{\bf K+Q}=0$ fixes the value $Q_{1X}=2\arccos{\left[({-2t + \mu})/({2 (t + 3 t')})\right]}$. The expansion of the  dispersion near Kohn points yields
\begin{eqnarray}
v&=&\sqrt{(6t'+\mu)(4t+6t'-\mu)(t^2+2tt'+3t'^2)}/(t+3t'),\\
m_x^{-1}&=&-\frac{4t^3+t^2(10t'-\mu)+4t'^2\mu +t t'(10t'+\mu)}{4(t^2+2t t'+3t'^2)},\nonumber\\
m_y^{-1}&=&-\frac{(t+4t')\mu -2t(2t+7t')}{4(t+3t')}.\nonumber
\end{eqnarray}
From the condition of the existence of Kohn points and presence of the maximum of the susceptibility (the signs of the inverse masses must be different), we obtain the following inequality for the values of $\mu$ and $t'$: 
\begin{equation}
\mu<\min\left[\frac{t(4t^2+10t t'+10t'^2)}{t^2-t t'-4t'^2},\frac{t(4t+14t')}{t+4t'}\right].
\label{Ineq_diag}
\end{equation}
The result (\ref{Ineq_diag}), together with the restriction of the chemical potential within the band $\mu>\min(E_X,E_L)$, where $E_X=4t+6t'$, $E_L=-6t'$ are the values of electronic dispersion (for $\mu=0$) at the points $X$ and $L$, respectively, is shown graphically in Fig. \ref{fig_reg}a.  
In the considered range $t'$ (except $t'\approx-t/3$) there is sufficiently wide range of chemical potentials $\mu$ where the Kohn points give the dominant contribution to the momentum dependence of the susceptibility.     

2) For the direction ${\mathbf Q}=(0,Q_{2X},Q_{2X})$ we have $Q_{2X}=2\arccos[({-2 t - 2 t'+ \mu})/({2 (t + 2 t')})]$,
\begin{eqnarray}
v&=&\sqrt{\frac{(-4t-6t'+\mu)(4t^2 + t\mu+2(t+\mu)t'+2t'^2)}{2t+4t'}},\\
m_x^{-1}&=&-2t',\nonumber\\
m_y^{-1}&=&\frac{4t^3+6 t^2 t'+4t'^3-t^2\mu-\mu^2t' }{4t^2+t(2t'+\mu)+2t'(2t'+\mu)}.\nonumber
\end{eqnarray}
From the condition for the existence of Kohn points and the presence of the anomaly (the signs of the inverse masses must be different), we obtain the following inequalities for the values of $\mu$ and $t'$ for $0< {t'}/{t} <{1}/{2}$: 
\begin{equation}
-\frac{2(2t^2+t t'+2t'^2)}{t+2t'}<\mu<\frac{-t^2+\sqrt{t^4+16t^3 t'+24t^2 t'^2+16t'^4}}{2t'}.
\end{equation}
However, Kohn anomalies can appear also in case of the same sign of the inverse masses, provided by $|m_x|\gg |m_y|$ or $|m_x|\ll |m_y|$.
The ranges of $t'$ and $\mu$ that fulfill these conditions are shown in Fig. \ref{fig_reg}b.

3) For the direction $(0,0,Q_{3X})$ we find 
$Q_{3X}=2\arccos[(-2t \pm \sqrt{4t^2 + (4t + \mu) t' - 2 t'^2})/(2 t')]$. 
We have $m_x=m_y$ by symmetry, 
and therefore in general case this direction does not lead to Kohn points that provide maximum of the susceptibility.
However, the corresponding mass 
\begin{eqnarray}
m_x^{-1}&=&\frac{2(t^2+t t' -2t'^2) \mp t\sqrt{4t^2+t'(4t-2t'+\mu)}}{2t'}
\end{eqnarray}
can be equal to zero for certain values of $\mu$, $t'$, providing a maximum of susceptibility.

4) For the direction $(Q_{4X},0,0)$ we have $Q_{4X}=2\arccos[(-4t-4 t'+\mu)/(2 t')]$,
\begin{eqnarray}
v&=&\sqrt{(4t+6t'-\mu)(-4t-2t'+\mu)},\nonumber\\
m_{x,y}^{-1}&=&-t-2t'\mp t\sqrt{\frac{-4t-2t'+\mu}{4t'}}.\label{m4X}
\end{eqnarray}
The corresponding regions of existence of Kohns anomalies in the susceptibility are shown in Fig. \ref{fig_reg}c.

\begin{table}[b]
\centering
$$\begin{array}{||l||c|c||c|c||}
\hline\hline
& \multicolumn{2}{c||}{t^{\prime }=0.3t} & \multicolumn{2}{c||}{%
t^{\prime }=-0.45t} \\ \hline\hline
(Q_{1X},Q_{1X},Q_{1X}) & -1.8t<\mu<3.73t & 0.55<n<1.71 & 1.3t<\mu<2.38t  & 0.93<n<1.38 \\ 
(0,Q_{2X},Q_{2X}) & -3.1t<\mu<3.07t & 0.38<n<1.45 & - & - \\ 
(0,0,Q_{3X}) & \multicolumn{1}{c|}{\mu \simeq 0^{**}} & 
\multicolumn{1}{c||}{n\simeq 0.91^{**}} & \multicolumn{1}{c|}{-} & 
\multicolumn{1}{c||}{-} \\ 
(Q_{4X},0,0) & - & - & 1.3t<\mu<3.08t & 0.93<n<2.00 \\ 
(Q_{3L},-Q_{3L},0) & \mu>-1.8t  & n>0.55 & 2.7t<\mu <3t & 1.60<n<1.96
\\ \hline\hline
\end{array}$$
\caption{
Ranges of chemical potentials and concentrations at which local susceptibility 
maxima, originating from Kohn points, are expected in symmetric directions at $t'=0.3t$ and $t'=-0.45t$.
For the values indicated by $^{**}$ both inverse masses $m^{-1}_{x,y}$ are close to zero.}
\end{table}

Similarly we can consider Kohn points located symmetrically with respect to the point $L$: for the wave vectors 
$(Q_{1L},Q_{1L},Q_{1L})$, $(Q_{2L},Q_{2L},0)$ and $(Q_{3L},0,0)$ we find 
\begin{eqnarray}
Q_{1L}&=&2\arccos[(6t+\mu)/(6t-6t')],\nonumber\\ 
Q_{2L}&=&2\arccos[(2t+2t'+\mu)/(2t-4t')],\nonumber\\ 
Q_{3L}&=&2\arccos[(-4t'-\mu)/(2t')],\nonumber
\end{eqnarray}
respectively; the masses $m_{x,y}$ are always of the same sign; 
for the wave vector $(Q_{4L},-Q_{4L},0)$ we have 
\begin{equation}
Q_{4L}=2\arccos[(2t-2t'-\mu)/(2t+4t')],\nonumber
\end{equation}
the corresponding conditions that the masses are opposite are shown graphically in Fig. \ref{fig_reg}d.

For Kohn points, which are symmetric with respect to the point $W$ and connected by the wave vector $(Q_{1W},0,0)$, we find
$Q_{1W}=2\arccos[(4t+4 t'-\mu)/(2 t')]$.
In this case, the masses for the point $(\pi-Q_{1W}/2,0,2\pi)$ will be the same as in the equation (\ref{m4X}) above
for points, symmetric with respect to the point $X$, characterized by the vector $(Q_{4X},0,0)$, 
and they interchange their values at the point $(\pi+Q_{1W}/2,0,2\pi)$. 

From these results it follows that for $t'>0$ Kohn points give a nonanalytic contribution to the susceptibility 
in a wide range of $-2t\lesssim \mu \lesssim 4t$, and for $t'<0$ in the range $t\lesssim \mu \lesssim 3t$.
Specific ranges of chemical potentials and concentrations at $t '= 0.3t$ and $t' = -0.45t$ are presented in Table. 1.

\subsection{The relation between the Kohn points and the momentum dependences of the susceptibility and
effect of Kohn points on quantum phase transitions}

The analysis of the previous subsection allows us to identify peaks of susceptibility, found numerically in Sec. 2.1,
with the contributions of various Kohn points, establishing the correspondence of these contributions with different types of magnetic order.

For $t'>0$ the evolution of maxima of the irreducible susceptibility in Fig. \ref{fig5} with decrease of the chemical potential corresponds to a change of the type of magnetic order FM $\,\,\rightarrow(Q_{W},Q_{W},Q_{W})$ $\rightarrow(Q_{1X},Q_{1X},Q_{1X})$ $\rightarrow(0,0,Q_{3X})$, with a possible narrow region of the phase $(0,Q_{2X},Q_{2X})$  between phases $(Q_{1X},Q_{1X},Q_{1X})$ and $(0,0,Q_{3X})$. The phase $(Q_{W},Q_{W},Q_{W})$, which was not considered in Section \ref{Sec:Kohn_points} because of very specific (disconnected) form of the Fermi surface, is present at large chemical potential near the upper edge of the band.
Further types of order $(Q_{1X},Q_{1X},Q_{1X})$ and $(0,0,Q_{3X})$, which occur with decrease of the chemical potential are determined by the vicinity of the point $X$  and characterized by the wave vectors $Q_{1X}$ and $Q_{3X}$, coinciding with those found in Section \ref{Sec:Kohn_points}. Although the sequence of the phases, determined in this paper, coincides with previously obtained \cite{Igoshev,Igoshev1}, there are some differences due to more accurate (analytical) determination of wave vectors in Section \ref{Sec:Kohn_points}. For example, the range of existence of the phase $(\pi,\pi,\pi)$, which is the limiting case of the $(Q_{1X},Q_{1X},Q_{1X})$ phase for $Q_{1X}\rightarrow \pi$, shrinks to the point 
$\mu=2t$.

On the other hand, at $t'<0$ the decrease of the chemical potential leads to the following sequence of dominating phases: FM$\,\,\,\,\rightarrow(0,0,Q_{W})\rightarrow(Q_{1X},Q_{1X},Q_{1X})$, the phase $(Q_{1X},Q_{1X},Q_{1X})$ is present at not too large $|t'/t|$. 
Although opening of Fermi surface ``window'' in the vicinity of point $L$ at $\mu=\mu_L=-6t'$ does not change the  symmetry of the wave vector, for which the global susceptibility maximum is achieved (by virtue of dominating contributions of the points $X$,$W$), the momentum dependence of $\chi_{\mathbf q}^0$ in the vicinity of $\mathbf{q} = 0$ changes significantly (see Fig.~\ref{fig6}b,c). 
In particular, for $\mu>\mu_L$ the susceptibility rather weakly depends on $\mathbf{q}$  at small $q$ (Fig.~\ref{fig6}c), analogously to the case of small $t'$ two-dimensional  $t$-$t'$ Hubbard model slightly above van Hove filling~\cite{Onufr,OurBook}, so that in this case the competition of ferromagnetic and incommensurate correlations occurs. 
Dominating type of magnetic ordering in this situation can be easily changed by correlations and/or peculiarities of the Fermi surface away from the $L$ point.  
At the same time, at $\mu<\mu_L$ the susceptibility has the pronounced minimum at $\mathbf{q}=0$ (Fig.~\ref{fig6}b) indicating the absence of ferromagnetic correlations, as well as dominating incommensurate correlations. 
The change of susceptibility momentum dependence when the Fermi surface crosses $L$ point is largely due to occurence of Kohn points connected by wave vectors $(Q_{1L},Q_{1L},Q_{1L})$ and $(Q_{2L},Q_{2L},0)$  in the vicinity of the $L$ point  at $\mu<\mu_L$. 
In all considered cases the susceptibility maximum is caused by hyperbolic Kohn points considered in Section 2.1. 
At finite temperatures the theory predicts  at quantum phase transition point critical exponents of the susceptibility $\chi_{\bf Q} \propto T^{-\gamma}$ and the correlation length $\xi \propto T^{-\nu}$ equal to $\gamma=\nu=1$, see Refs.~\cite{Rice,Katanin}.

\section{The antiferromagnetism of chromium and the two-band model} 

\begin{figure}
\centering
\includegraphics[width=0.4\textwidth]{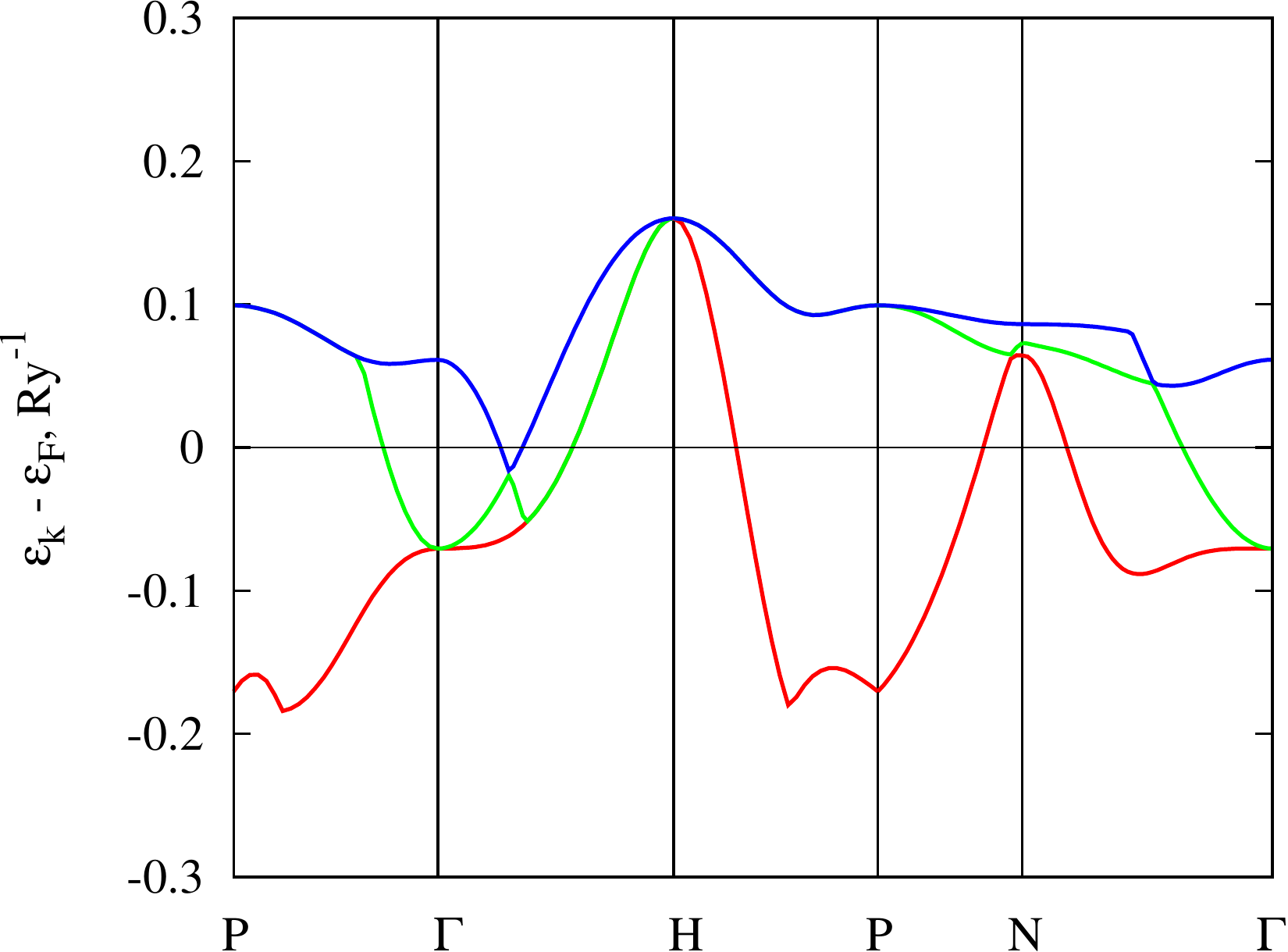}
\includegraphics[width=0.36\textwidth,
trim=0 -20 40 40
]{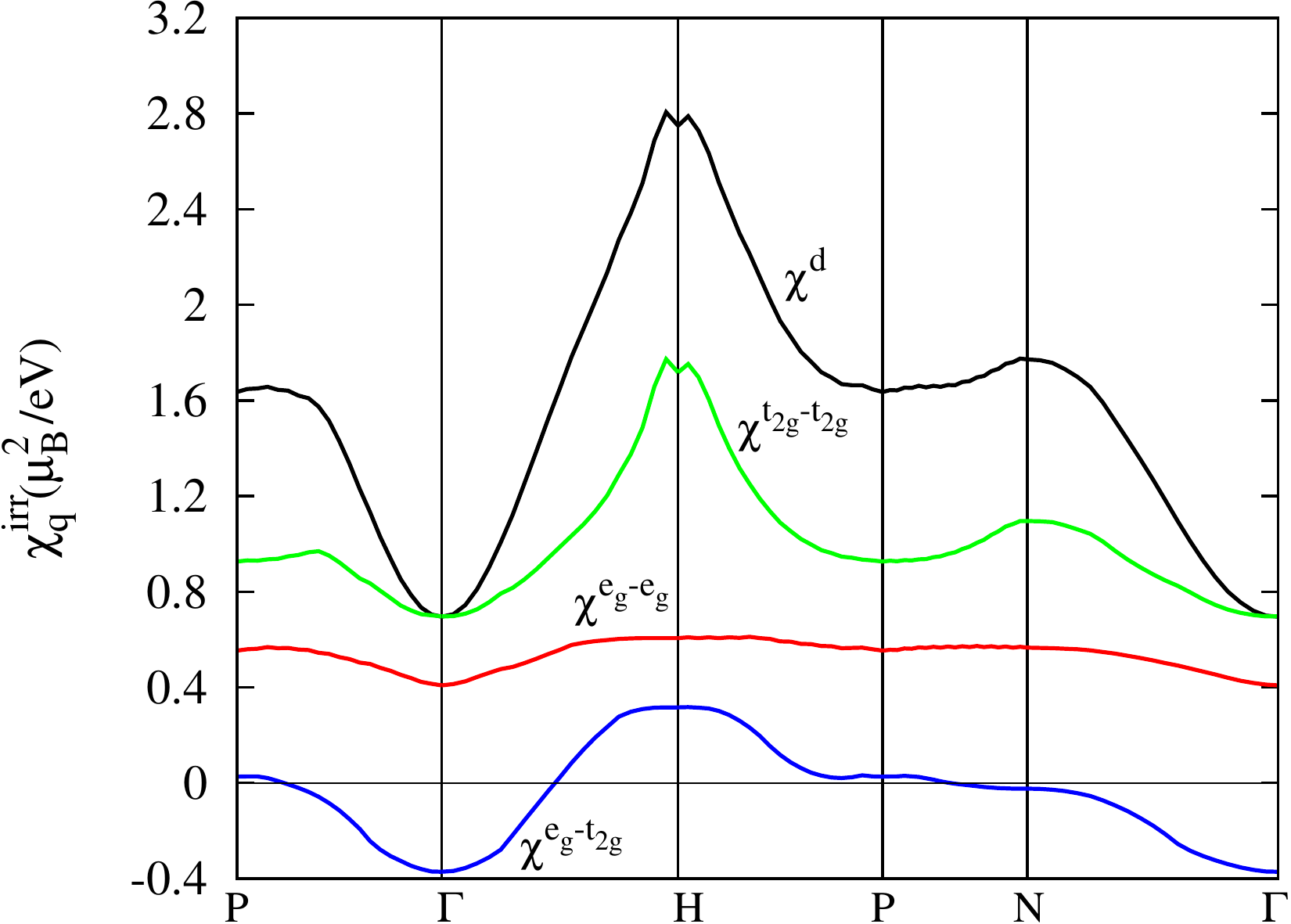}
\caption{The band structure in the vicinity of Fermi level and the momentum dependence of the susceptibility of paramagnetic chromium along symmetric directions, obtained within the first-principle calculations}\label{fig_bcc_Cr1}
\end{figure}

To study the nature of antiferromagnetism of chromium we consider first the results of ab initio calculations within the local density approximation~\cite{Jones_Gunnarsson_RMP_1989} using tight-binding linear muffin-tin orbital atomic spheres 
framework~(LDA~TB-LMTO-ASA)~\cite{Andersen_Jepsen_PRL_1984}.
The von Barth-Hedin local exchange-correlation potential has been used~\cite{Barth_Hedin_1972}; 
we choose the lattice parameter $a=2.8845$~\AA\ and the mesh 40$\times$40$\times$40  in the reciprocal space. 
The chromium band structure and momentum dependence of the susceptibility, together with partial contributions of differtent $d$ orbitals, are presented in Fig.~\ref{fig_bcc_Cr1}. 
The susceptibility has a maximum at the incommensurate wave vector $(0,0,Q)$ in $\Gamma-H$ direction with $Q$ close to $2\pi$. 

\begin{figure}[t]
\centering
\includegraphics[width=0.4\textwidth]{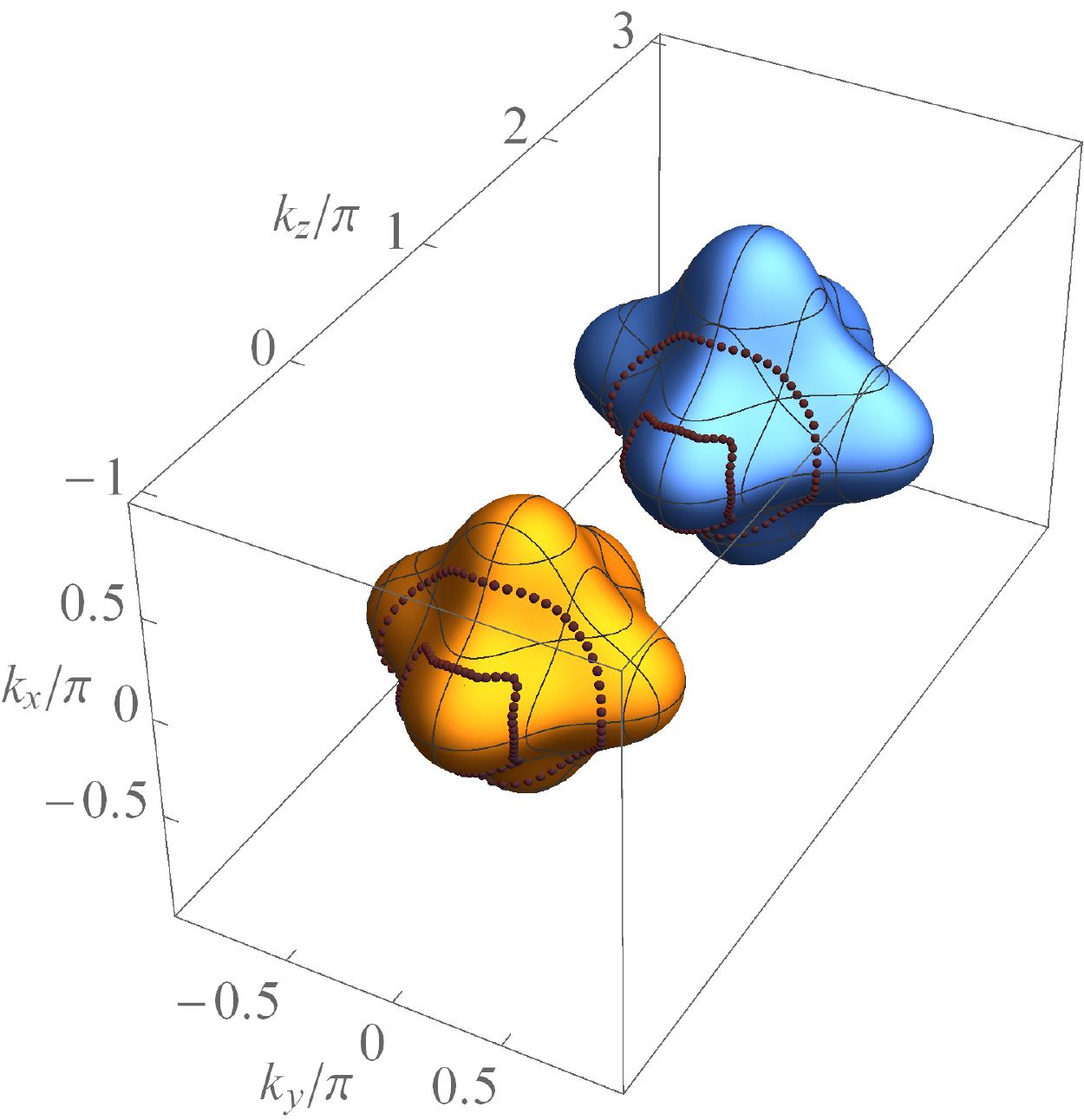}
\caption{Two-band model Fermi surfaces for values of the parameters, specified in Table~\ref{tab1}. 
The positions of Kohn points  (see text) are shown by dots.}\label{fig_2band}
\end{figure}

To explain the obtained momentum dependence of the susceptibility and establish the connection with Kohn anomalies, we consider a simple two-band model, qualitatively describing the chromium Fermi surfaces, $\hat H=\sum_{{\bf k},\sigma}E_m({\bf k})\hat c_{{\bf k} m \sigma}^+\hat c_{{\bf k} m \sigma}$, $m=1,2$ being the band index.  
The corresponding dispersion is
\begin{eqnarray}
E_{\mathrm{\mathbf{k}}}^{(m)}=&-&8t_{1m}\cos{\frac {k_x}{2}}\cos{\frac {k_y}{2}}\cos{\frac {k_z}{2}}\notag\\&-&2t_{2m}\left(\cos{k_x}+\cos{k_y}+\cos{k_z}\right)\notag\\&-&4t_{3m}\left(\cos{{k_x}}\cos{{k_y}}+\cos{ {k_x}}\cos{ {k_z}}+\cos{ {k_y}}\cos{ {k_z}}\right)+E_{0m},
\label{TwoBand}
\end{eqnarray}
where $t_{1m}, t_{2m}, t_{3m}$ are nearest-, next-nearest and next-next-nearest neighbour hopping integrals of body centered cubic lattice for the first and second bands respectively, $E_{0m}$ are energy levels of the bands.  
Hopping integrals and energy levels are determined by the coincidence of Fermi surface points along symmetric directions, obtained within the first-principle calculations and the model~(\ref{TwoBand}).  
The resulting parameter values are presented in Table~\ref{TableHoppingsCr}, 
the corresponding Fermi surfaces are shown in Fig.~\ref{fig_2band}. 

\begin{table}[h]
$$\begin{array}{|c|r|r|r|}
\hline
m & t_{2m}/t_{1m} & t_{3m}/t_{1m} & E_{0m}/t_{1m} \\
\hline
1 & 0.91 & -0.59 & 6.61 \\
\hline
2 & -1.52 &  0.80 & -7.70 \\
\hline
\end{array}$$
\caption{\label{TableHoppingsCr}Next- and next-next-nearest neighbour hopping integrals and energy levels of the bands in units of nearest neighbour hopping integrals}\label{tab1}
\end{table}

\begin{figure}[b]
\centering
(a)\includegraphics[width=0.43\textwidth]{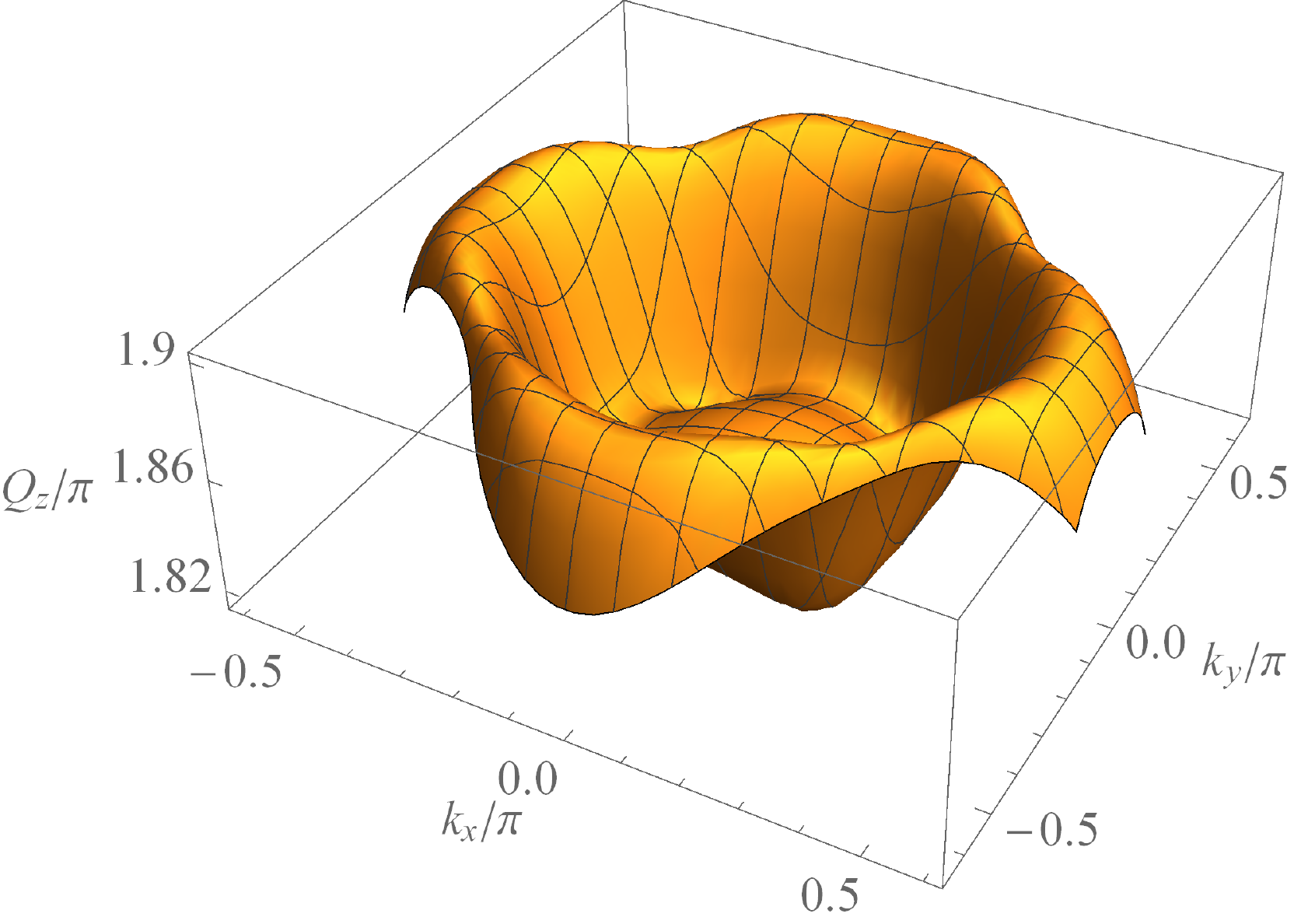}
\hspace{0.5cm}(b)\includegraphics[width=0.4\textwidth]{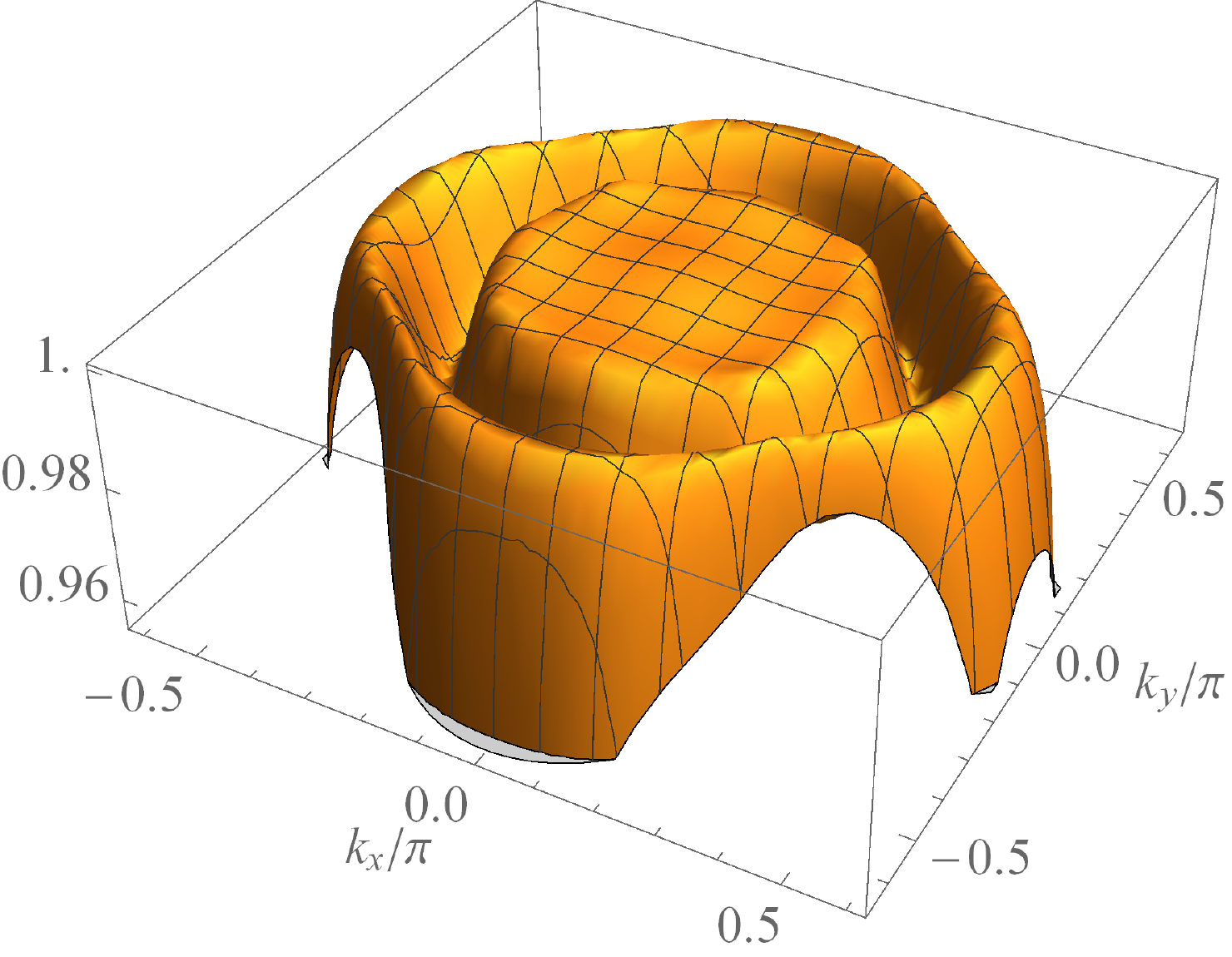}
\caption{(a) The length of the wave vector $Q_z(k_x,k_y)$, connecting Fermi surface points $(k_x,k_y,k_z)$ and $(k_x,k_y,k_z+Q_z)$ of the first and the second bands; (b) Values of $-\cos({\mathbf{v}_1(\mathbf{k}), \mathbf{v}_2(\mathbf{k+Q})})$ for different points of the Fermi surface}\label{figCOS}
\end{figure}

It is worthwhile to note that the considered model does not assume approximate nesting between different sheets of Fermi surface and in this respect it is more realistic than models of antiferromagnetism of chromium considered earlier, see, {e.g.},~ Refs.~\cite{Lomer,Shibatani,Dzyaloshinskii,Machida}.   
At the same time, the absence of nesting in our model even improves qualitative applicability  of random phase approximation, in comparison to  the case of perfect nesting, discussed previously in Refs.~\cite{Dzyaloshinskii,Dzyaloshinskii1}.

The dominant contribution to the susceptibility in the present model is expected from Kohn points on different sheets of the Fermi surface: electron-like sheet closed around the point $\Gamma$ and hole-like sheet closed around the point $H=(0,0,2\pi)$. 
The  expansion of the spectrum in the vicinity of Kohn points on different sheets yields different magnitudes of Fermi velocity and inverse masses. 
To find the Kohn points we assume that the vector ${\bf Q}=(0,0,Q_z)$ is parallel to one of coordinate axis (which corresponds to the results of ab initio calculations and experimentally observed vector ${\bf Q}$) and enforce the conditions that (a) the points ${\bf K}$ and ${\bf K+Q}$ belong to Fermi surfaces of first and second band respectively, (b) the corresponding Fermi velocities ${\bf v}_1({\bf K})$ and ${\bf v}_2({\bf K+Q})$ are antiparallel. 
Due to the symmetry it is sufficient to consider only the part $K_z<0$ of the Fermi surface of the first sheet, which can be parametrized by coordinates $K_x,K_y$ of point ${\bf K}$. 
The quantity $Q_z(K_x,K_y)$, determined by the condition (a), is therefore also a single-valued function of these two coordinates of Kohn point on the first sheet, the result of calculation of this function is plotted in Fig.~\ref{figCOS}a.  
To verify the condition (b), we show the values of 
$-\cos{(\mathrm{\mathbf{v}_1({\bf K})}, \mathrm{\mathbf{v}_2({\bf K+Q})})}=-\mathbf{v}_1({\bf K})\cdot \mathbf{v}_2({\bf K+Q})/({v}_1({\bf K}) {v}_2({\bf K+Q}))$ in Fig.~ \ref{figCOS}b. 
One can see that Kohn points form two lines in the reciprocal space, characterized by values 
$Q_{1z}\approx 1.83\pi$ and $Q_{2z}\approx 1.90\pi$. 
Moreover, the inner line of Kohn points with $Q_z=Q_{1z}$ is a boundary of a region, in which Fermi velocities on different sheets are almost antiparallel. 
This region is, however, characterized by substantial dependence of $Q_z$ on $K_{x,y}$ and, therefore, does not  substantially contribute to the momentum dependence of the susceptibility, which is confirmed by a~numerical calculation of the susceptibility, see below. 
The positions of Kohn point lines on the Fermi surfaces  are shown in Fig.~\ref{fig_2band}. 

To obtain the momentum dependence of the susceptibility, we parametrize the position of Kohn points ${\bf K}_i(\varphi)$ on each line (enumerated by $i=1,2$) by an angle $\varphi$ and introduce local rotated coordinate system in reciprocal space such that $k'_{zi}$ axis  is directed along the difference of Fermi velocities ${\mathbf{v}_{1}}({\bf K}_i(\varphi))-{\mathbf{v}_{2}({\bf K}_i(\varphi)+{\bf Q}_i)}$ at the corresponding Kohn points, while $k'_{xi}$ axis is directed perpendicular to $k'_{zi}$ and tangentially to $i$-th line of  Kohn points on the first sheet. 
Moreover, on Kohn point lines, where Fermi velocities of the two sheets are almost opposite (see Fig.~\ref{figCOS}b), $k'_{xi}$ axis lies in a tangential plane to the Fermi surface. 

The corresponding expansion of the spectrum in the vicinity of Kohn points has a form
\begin{eqnarray}
\label{E2}
E_{{\mathbf{K}_i(\varphi)+\mathbf{k}}}^{(1)}&\simeq& v_{1i}(\varphi)k'_{zi}+\frac {(k'_{xi})^2}{2m_{1i}(\varphi)},\notag\\
E_{{\mathbf{K}_i(\varphi)+\mathbf{Q}_i+\mathbf{k}}}^{(2)}&\simeq& -v_{2i}(\varphi)k'_{zi}+\frac {(k'_{xi})^2}{2m_{2i}(\varphi)},
\end{eqnarray}
where $v_{1i}(\varphi)=|\mathbf{v}_1({\bf K}_i(\varphi))|$, $v_{2i}(\varphi)=| \mathbf{v}_2({\bf K}_i(\varphi)+{\bf Q}_i)|$, and we neglect weak dependence ${\mathbf{Q}_i(\varphi)}$, i.~e. we assume that lines of Kohn points are approximately parallel.  
Due to the symmetry of the Fermi surfaces one can restrict oneself by angles $\varphi \in[0,\pi/4]$. 
To reduce the dispersion to standard form~(\ref{Eexp}) we perform change of variables
\begin{eqnarray}
k'_{zi}&\to& k'_{zi}-\frac{(k'_{xi})^2}{2(v_{1i}(\varphi)+v_{2i}(\varphi))}\left[\frac {1}{m_{1i}(\varphi)}-\frac {1}{m_{2i}(\varphi)}\right].
\end{eqnarray}
Written in new variables, the spectrum has the form
\begin{eqnarray}
E_{
{\bf K}_i(\varphi)+{\mathbf{k}}}^{(1)}&\simeq& v_{1i}(\varphi)k'_{zi}+\frac{(k'_{xi}) ^2}{2m_{si}(\varphi)},\notag\\
E_{
{\bf K}_i(\varphi)+{\bf Q}_i+{\bf k}}^{(2)}&\simeq& -v_{2i}(\varphi)k'_{zi}+\frac{(k'_{xi})^2}{2m_{si}(\varphi)},
\end{eqnarray}
where
\begin{eqnarray}
\label{Rm1}
\frac{1}{m_{si}(\varphi)}&=&\frac{1}{v_{1i}(\varphi)+v_{2i}(\varphi)}\left[\frac{v_{1i}(\varphi)}{m_{2i}(\varphi)}+\frac{v_{2i}(\varphi)}{m_{1i}(\varphi)}\right].
\end{eqnarray}

\begin{figure}[t]
\centering
\subfloat[]{
\label{figMx1}
\includegraphics[width=0.4\textwidth]{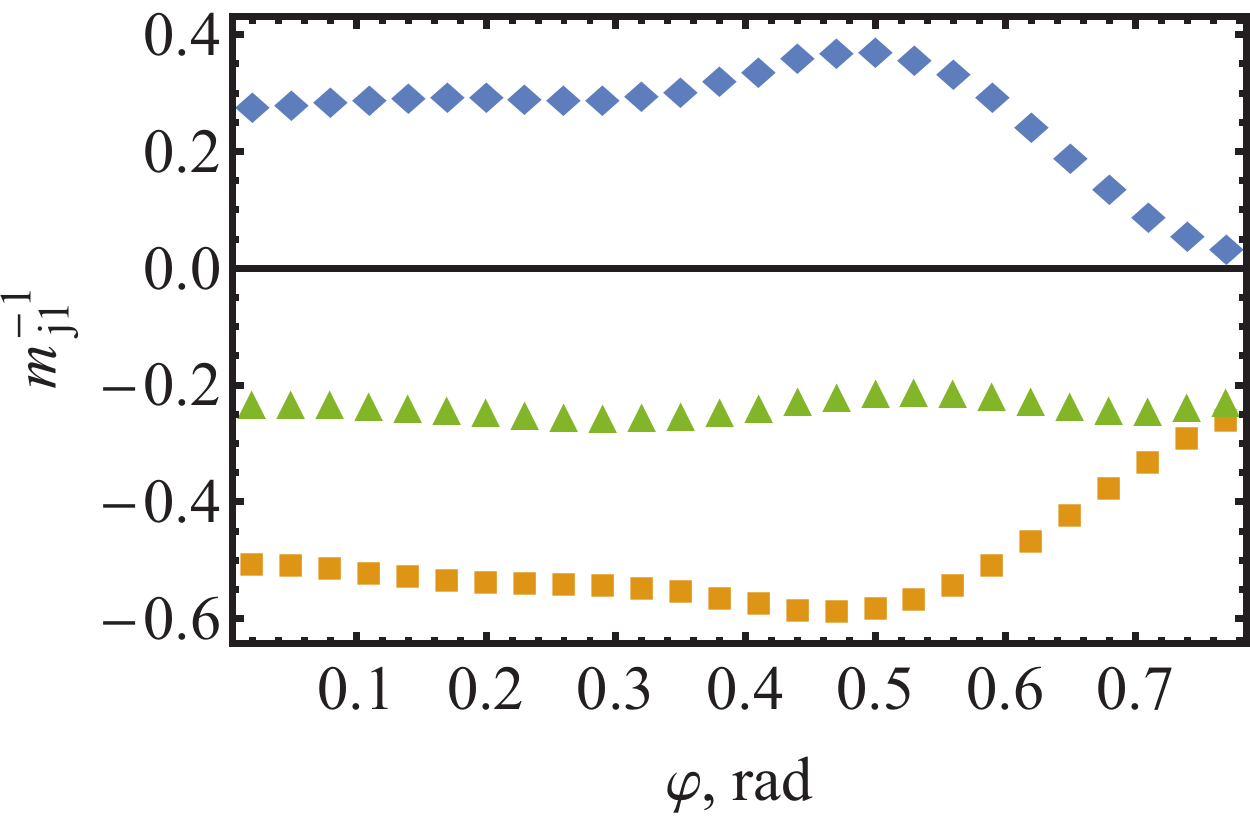}}
\subfloat[]{
\label{figMy1}
\includegraphics[width=0.4\textwidth]{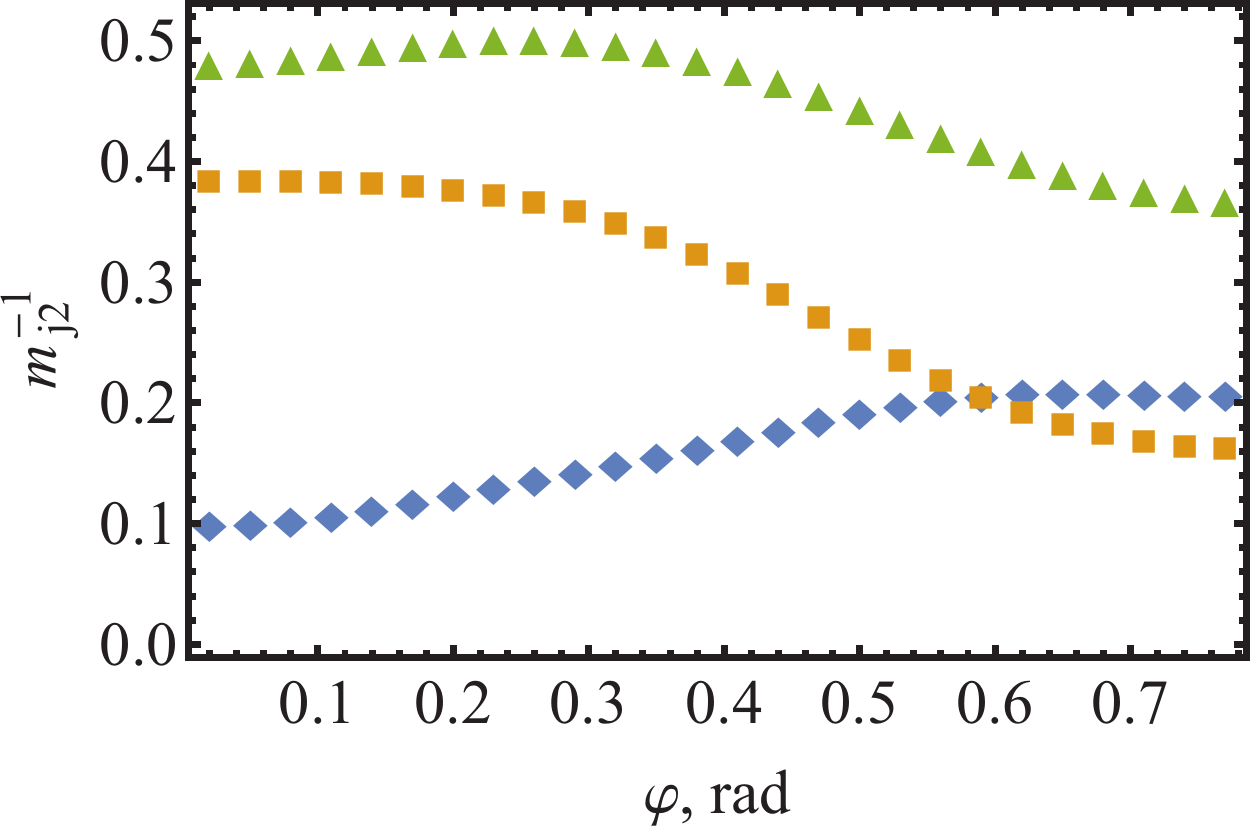}}
\caption{The dependence of inverse masses  $m_{ji}^{-1}(\varphi)$ for inner, $i=1$ (a) and outer, $i=2$ (b) Kohn point lines on an angle $\varphi$ counted from the direction $\Gamma X$. Rhombs correspond to $m^{-1}_{1i}$, squares -- to $m^{-1}_{2i}$, and triangles -- to the resulting $m^{-1}_{si}$, calculated using Eq.~(\ref{Rm1})}\label{figM1}
\end{figure}

The dependences of inverse masses  $m_{ji}(\varphi)$, $m_{si}(\varphi)$ ($i,j=1,2$) are shown in Fig.~\ref{figM1}. 
The signs of resulting masses $m_{si}(\varphi)$ do not change with $\varphi$ but opposite for the two Kohn point lines. 
The contribution of these points to interband part of the susceptibility at $T=0$ can be calculated analogously to Refs.~\cite{Rice,Stern,Holder2012,Metzner,Katanin},
\begin{eqnarray}
\chi_{\mathrm{\mathbf{q+Q}}}^{{\rm ib}, 0}&\equiv&
-\sum_{\mathrm{\mathbf{k}}}\frac {f(E_{\mathrm{\mathbf{k}}}^{(1)})-f(E_{\mathrm{\mathbf{k+q+Q}}}^{(2)})}{E_{\mathrm{\mathbf{k}}}^{(1)}-E_{\mathrm{\mathbf{k+q+Q}}}^{(2)}}\notag\\&\simeq&
\chi_{\mathrm{\mathbf{Q}}}^{{\rm ib}, 0}-\frac{4}{\pi^2}\sum_{i=1,2}\int_0^{\pi/4} \frac{d\varphi}{D_i(\varphi)}\frac{1}{v_{1i}(\varphi)+v_{2i}(\varphi)}
\left\{
\begin{array}{ll}
\sqrt[]{|V_{si}(\varphi)q'_{zi}(\varphi)}|, & m_{si}(\varphi)q'_{zi}(\varphi) <0,\\
{
V_{si}(\varphi)}q'_{zi}(\varphi)/(\pi {\Lambda}), & m_{si}(\varphi)q'_{zi}(\varphi) >0,
\end{array}
\right.
\label{chi12_cr}
\end{eqnarray}
where
\begin{eqnarray}
V_{si}(\varphi)&=&\frac{2m_{si}(\varphi)v_{1i}(\varphi)v_{2i}(\varphi)}{v_{1i}(\varphi)+v_{2i}(\varphi)},\\
D_i(\varphi)&=&\left|\frac{\partial(k_x,k_y,k_z)}{\partial(\varphi,k'_{xi},k'_{zi})}\right|_{k'_{xi}=k'_{zi}=0},
\end{eqnarray}
$q'_{zi}(\varphi)$ being the projection of vector $\bf q$ onto the axis $k'_{zi}$, $\Lambda \sim 1$ is the cutoff parameter in the reciprocal space. 

\begin{figure}[h]
\centering
\includegraphics[width=0.45\textwidth]
{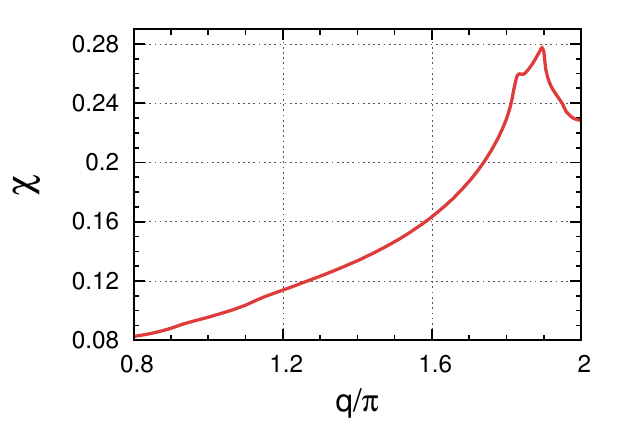}
\caption{Interband contribution to the momentum dependence of the magnetic susceptibility of two-band model at $T=0$ in the direction $(0,0,q)$.
}
\label{figHi_Cr}
\end{figure}
 
Therefore, in this case the square-root non-analytic momentum dependence of the susceptibility is expected at wave vectors close to $\mathbf{Q}_{1,2}$ and connecting Kohn point lines on different sheets. 
This conclusion is confirmed by the calculation of momentum dependence of interband contribution to the magnetic susceptibility within the two-band model: the result of numerical calculation of the integral in the first line of Eq.~(\ref{chi12_cr}) is presented in Fig.~\ref{figHi_Cr}, showing that the local non-analytic maxima of magnetic susceptibility are exactly positioned at the wave vectors $\mathbf{Q}_{1,2}$. 
Moreover, the contribution of outer Kohn point line with wave vector ${\bf Q}_2$, which is close to experimentally determined wave vector of spin density wave in chromium~\cite{Cr,Cr_dop3}, appears to be dominating. 
The analysis of the susceptibility at finite temperatures in the considered case of cylindrical Kohn points (under the assumption that the band structure is not substantially changed with pressure or doping)
predicts the critical exponents $\gamma=\nu=1$ of susceptibility and correlation length at the quantum phase transition point  with weak logarithmic corrections, see Refs.~\cite{Holder2012,Katanin}.

\section{Conclusion}

In this paper we have investigated the effect of Kohn points in electronic spectrum  of three-dimensional systems on magnetic properties within a one-band model with hopping between the nearest- and next-nearest neighbors on fcc lattice and two-band model, containing hopping within three coordination spheres on bcc lattice and modeling some features of the electronic dispersion of chromium.

For fcc lattice, we have investigated the effect of Kohn hyperbolic points on the magnetic susceptibility in a wide range of chemical potentials (concentrations) and determined the ranges of chemical potentials for symmetric directions, in which the effect of these points is expected. The obtained results are confirmed by the numerical analysis of the susceptibility of non-interacting electrons on fcc lattice. Near quantum phase transitions the studied Kohn points lead to a temperature dependence of the susceptibility and the correlation length $\chi_{\bf Q} \propto \xi \propto 1/T$. 

We have additionally investigated the effect of the opening of the window of Fermi surface near the point $L$, which may be important for explaining the properties of ZrZn$_2$. It is shown that when the point $L$ of the Brillouin zone belongs to the filled states ($E_L$<0, the Fermi surface window is open near $L$), there is a competition of ferromagnetic and incommensurable correlations. Closing this window ($E_L>0$) leads to a drastic change of the momentum dependence of the magnetic susceptibility and substantial increase of the contribution of incommensurate correlations.

On the basis of the analysis of the two-band model of  chromium, it is shown that two lines of Kohn points may be present in this substance, corresponding to close values of antiferromagnetic wave vector $(0,0,Q_z)$ with $Q_{1z}\approx 1.83\pi$ and $Q_{2z}\approx 1.90\pi$ in units of the inverse lattice constant. These lines of Kohn points lead to a one-sided nonanalytic (square root) momentum dependence of the susceptibility at $T=0$, the contribution of the outer line of Kohn points with $Q_z=Q_{2z}$, which is close to the experimentally measured spin-density wave vector in chromium, dominates. At finite temperatures, temperature dependences of the susceptibility and the correlation length $\chi_{\bf Q}\propto \xi \propto 1/T$ with weak logarithmic corrections are expected near quantum phase transition.

The described approaches to study the contribution of Kohn points can further use realistic band stucture, obtained within ab initio investigations in the framework of the density functional method and the dynamic mean-field theory, which will allows one to investigate the effect of Kohn points in real substances. The application of the dynamic mean-field theory \cite{DMFT} and dynamical vertex approximation \cite{DGA1,DGA2} will also allow one to investigate the effect of electronic interaction (including the non-local interactions) beyond random phase approximation. Although previous study of one-band model \cite{Katanin} showed that the electron-electron interaction does not change qualitatively the effect of Kohn  anomalies near quantum phase transitions, this problem requires more detailed study in future.

A detailed analysis of the available experimental data on magnetic properties of weak ferro- and antiferromagnets near quantum phase transitions will allow to analyze possible deviations from the predictions of the Hertz-Moriya-Millis theory caused by the effect of Kohn anomalies.

{ \it Acknowledgments} A. A. Katanin thanks T. Sch\"afer, A. Toschi, K. Held, and W. Metzner for valuable discussions of the effect  of Kohn anomalies in strongly correlated systems. The work was carried out within the state assignment of FASO of Russia (theme “Electron” 01201463326), and supported in part by grant from the Russian Foundation for Basic Research 17-02-00942a and project of the Ural Branch of the Russian Academy of Sciences 15-8-2-9.


\end{document}